# Super-Resolution imaging of plasmonic Near-fields:

# Overcoming Emitter Mislocalizations

*Yuting Miao[1], Robert C. Boutelle[2], Anastasia Blake[3], Vigneshwaran Chandrasekaran[3], Jennifer Hollingsworth[3], Shimon Weiss[1,4,5,6,*]*

[1]Department of Chemistry and Biochemistry, University of California, Los Angeles 90095, USA

[2]National Institute of Standards and Technology, 100 Bureau Drive, Gaithersburg, MD 20899, USA

[3]Los Alamos National Laboratory, Casa Grande Dr, Los Alamos, NM 87544, USA

[4]Department of Physiology, University of California, Los Angeles 90095, USA

[5]California NanoSystems Institute, University of California, Los Angeles 90095, USA

[5]Department of Physics, Institute for Nanotechnology and Advanced Materials, Bar-Ilan University, Ramat-Gan 52900, Israel

AUTHOR INFORMATION

**Corresponding Author**

* sweiss@chem.ucla.edu

**ABSTRACT**

Plasmonic nano-objects have shown great potential in enhancing biological and chemical sensing, light harvesting and energy transfer, and optical and quantum computing to name a few. Therefore, an extensive effort has been vested in optimizing plasmonic systems and exploiting their field enhancement properties. Super-resolution imaging with quantum dots (QDs) is a promising method to probe plasmonic near-fields, but is hindered by the distortion of the emission intensity and radiation pattern. Here we investigate the interaction between QDs and 'L-shaped' gold nanoantennas, and demonstrate both theoretically and experimentally that this strong interaction can induce polarization-dependent modifications to the apparent QD emission intensity, polarization and localization. Based on FDTD simulations and polarization-modulated single-molecule microscopy, we show that the displacement of the emitter's localization is due to the interference between the emitter and the induced dipole and can be up to 100 nm. We also discovered that the emission polarization can rotate towards the symmetry axis or one arm of the L-shape because of the scattering. Our results could assist in paving a pathway for higher precision plasmonic near-field mapping and its underlying applications.

**KEYWORDS**      Plasmonics, super-resolution, near-field, quantum dot.

With the development of plasmonics-based devices, there is a growing need for detecting and characterizing plasmonic effects in extended nanosystems. Due to their ability to concentrate light to a small dimension and create enormous local-field enhancement, nanoscale plasmonic devices have provided novel ways of controlling light and have shown great potential in broad applications, including enhanced chemical sensing [1-2], bio-sensing [3-5] and high-resolution bioimaging [6-7]. In addition, integrated nanophotonic circuits combing the plasmonic and optical effects have shown great promise in manipulating optical information [8-10]. Thus, to better control and utilize plasmonic near-field effects, a thorough understanding of the relationship between plasmonic structures and their local fields is crucial for optimizing these devices' performance. Compared to imaging methods like near-field scanning optical microscopy (NSOM) [11-13] or electron energy loss spectroscopy [14-15], super-resolution imaging has become more popular due to its ability to break the diffraction limit, image under ambient conditions, and provide high throughput imaging [16-18].

Recent studies have applied super-resolution fluorescent imaging with single emitters (e.g., dyes or quantum dots (QDs)) to probe plasmonic systems. In these studies, emitted fluorescence intensity from the emitter is used as a far-field reporter of the plasmonic near-field intensity [19-22]. However, due to the strong electromagnetic interaction between emitters and nearby plasmonic nanostructures, this technique is hindered by a complex mechanism. One major factor is the formation of a distorted point spread function (PSF) [23-24]. Since an accurate super-resolution localization relies heavily on a stable, well-characterized PSF, this distortion introduces error in the field intensity mapping [25-27]. It has been demonstrated that the fitted centroid position can move away from the actual emitter location when the probe emission is coupled to plasmonic antennas (e.g., nanorods [28], nanowires [29-31], and Yagi-Uda antennas [32]). This 'mislocalization' phenomenon may be originated from (1) super-position of the molecule emission and scattered radiation from the plasmonic interface, (2) molecule emission interference with the induced image dipole, and (3) near-field coupling of the molecule to the antenna [24,26,30,33]. These plasmon-induced interactions may redirect the single-molecule fluorescence polarization (mispolarization) as well [28,34-36]. Moreover, because of effects like fluorescence enhancement and quenching, the intensity of fluorescence can vary non-monotonically with the field intensity, especially when the emitter is too close (<30 nm) to the nanoantenna [19].

Despite intense interest and research activity, the interaction between nanoantennas and nearby molecules, as well as its influence on emitters' mislocalization and mispolarization, have not been completely understood. In this work, we select quantum dots (QDs) as the probe due to its degenerate excitation and emission dipole moments, and match their emission wavelength to be either on or off the plasmon resonance mode. We show theoretically and experimentally that this strong interaction can induce polarization-dependent changes to both apparent emission intensity and position. We extend previous studies to plasmonic nanoantennas with more complicated structural features, and demonstrate that shifts in the apparent emission localization and polarization are affected by a combination of factors under different conditions (e.g., emitter dipole location and orientation). Moreover, after optical measurements, we add a 'post mortem' scanning electron microscope (SEM) step to unveil the real position of QDs near plasmonic structures, and find out that the mislocalization can be up to ~100 nm and mispolarization can be up to ~30°. We isolate the effects from nanoantenna's two orthogonal structural features and elucidate the mechanisms behind emission localization and polarization modification. We also propose a new sample fabrication method that implements dip-pen nanolithography (DPN) [37] to achieve single-molecule deposition close to nanoantennas with high precision and throughput. Our work can provide a cost-effective, high-accuracy solution for better super-resolved mapping of plasmonics near-fields. It opens doors for optimized and controlled plasmonic devices with great potential in a wide area of applications.

*Tunable Nanoantenna Sample Design*. Previously, published works developed models on specific structures like nanowires and nanorods, and used simulation results to correct the mislocalization detected in the experiment. To expand on these previous models and on the 'one-dimensional' antenna structures, and in order to gain better understanding on how near-field coupling and far-field interference would affect the degree of QD mislocalization and miapolarization in a more complex (2D) model system, we performed extensive finite domain time difference (FDTD) calculations to determine the ideal antenna structure for our study (Figure 1a). With the help of these calculations, we converged to a 2D antenna design of an L-shaped gold nanostructure with a symmetry axis and a sharp corner, which are features that have not been carefully studied as of yet. The emitter at the inner corner of the L-shape can have simultaneous interactions with two arms. In this way, the coupling strength can be tuned by controlling the relative distance of QDs to each of the arm. Since the scattering from the antenna interface is one significant contribution to the QD image distortion, the relationship between the L-shape dimensions and their optical response is explored. L-shaped nanoantenna with various dimensions exhibit different scattering spectrum (Figure 1c), and QDs with varying emission wavelengths can be selected to be on- or off-resonance with the plasmon resonance mode. As for our experimental setting, the structure with a configuration of 60 nm height, 250 nm arm length, and 100 nm arm width has a much stronger scattering at 800nm than 600nm, while vice versa for the (H 60nm, L 200nm, W 50nm) configuration. Thus, by tuning the L-shape dimensions and choosing different QDs to match/mismatch the plasmon resonance, we can explore the scattering effect on the QD localization accuracy. For instance, 800-nm emission QDs are excited at a wavelength off the plasmon resonance (642 nm) to excite the emitter only and limit the antenna background. In this way, we avoid fluorescence absorption enhancement and isolate the effect in QD emission for polarization-modulated studies.

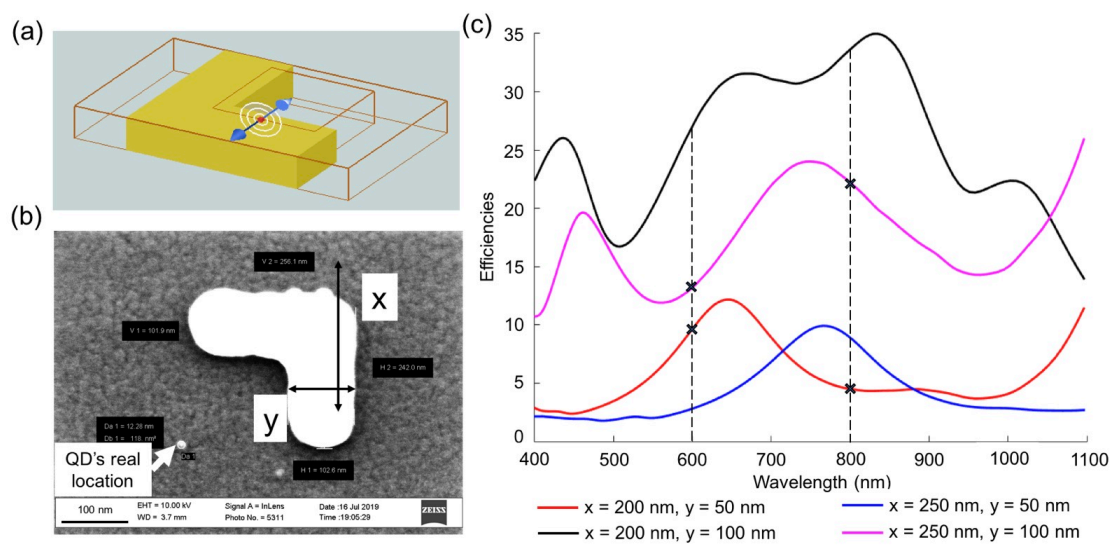

**Figure 1: L-shape dimensions and scattering cross-sections.** (a) Design for FDTD simulations with an L-shaped gold nanoantenna and a dipole source positioned nearby. (b) SEM image of the fabricated nanoantenna and QD. QD's real location measured from SEM can be fed-back to the simulation as input. (c) L-shaped Au nanostructures with different dimensions vary in the scattering cross-section at different wavelength. For a nanoantenna with a given dimensions, 800/600 nm emission QDs can be scattered strongly/weakly or weakly/strongly by the designed structure.

To construct a system with plasmonic structures with different dimensions and single-molecule QDs close to them, we employ a two-step lithography method (Figure 2). The first step defines the L-shaped plasmonic structure, and the second step sets the QD deposition boundaries. More specifically, nanoantennas with varying dimensions were first patterned onto an indium tin oxide (ITO)-coated glass coverslip with electron beam (e-beam) lithography. This conductive and transparent substrate prepares the sample for both optical and SEM measurements. The second lithographic step defines a pattern that prevents QD from accessing other areas of the coverslip surface, allowing QDs to deposit only near the inner corner of the antenna. Then, a drop of nano-molar QD solution is placed onto the coverslip to produce a sparse distribution of the probes. QDs can move around the whole substrate surface through Brownian motions, and those that reach the pre-defined area can attach to the surface through chemical functionalization. The carboxylic acid functional group in QD ligands can interact with the ITO surface through a combination of weak electrostatic, hydrogen-bonding, and covalent bonding between the carboxylate and the indium defect sites [38]. By tuning the QD concentration and the area during the second lithographic step, we ensure that an individual unit only has a single plasmonic nanoantenna and a single QD nearby. This single QD and antenna pair is further verified by the SEM image (Figure 1b). The difference between the dimensions of the fabricated and simulated structures is negligible. However, this sample fabrication method requires precise alignment between two e-beam lithographic steps, making the success rate / yield of fabrication low. We also combine dip-pen nanolithography (DPN) method [37,41] together with E-beam lithography to directly 'write' single QDs with higher precision (Figure S1). Compared to points accumulation for imaging in nanoscale topography (PAINT) methods that have been used for previous research which relies on the absorption and release of freely-diffusing probes at random locations on the sample surface [28], our method immobilizes QDs to sites close to the antenna with selective binding and can be only removed though specific washing steps. Instead of counting on the probe to stay in the field-of-view (FOV) long enough to emit enough photons (especially at low excitation power), we can keep QDs stationary with a well-defined dipole moment orientation throughout the whole acquisition for better PSF fitting and localization with high precision at multiple imaging conditions (e.g., excitation and emission polarization). The inter-antenna spacing is designed to be 5 $\mu$m to avoid interactions between plasmonic nanoantennas while keeping a reasonable measurement throughput. For each wide-field imaging run, images from up to 25 pairs of single antenna and QD can be collected. To ensure that emitters at different locations of the FOV share the same excitation intensity and polarization, the optical setup is first calibrated with free QDs at different polarizations. More details about the sample fabrication are provided in the Methods section.

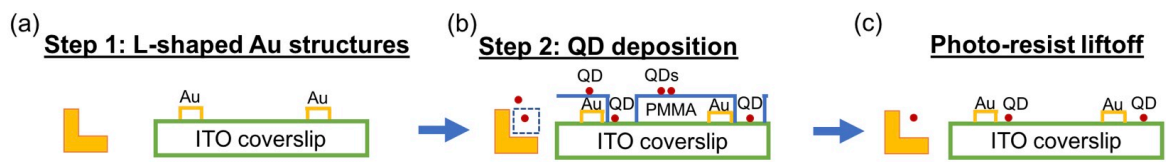

**Figure 2: 2-step e-beam lithography sample fabrication procedures.** (a) For the first electron beam lithography step, L-shapes with different dimensions are patterned. (b) The pattern of the second lithography layer is aligned to the first layer and is composed of squares that are accessible to QDs. After removing the exposed photoresist, the ITO coverslip was soaked in the diluted QD solution. The carboxyl group in QD ligands can then bind to ITO and stay static on the coverslip surface. (c) The photoresist lift-off step washes the unexposed photoresist off the coverslip, taking away QDs that are not bound to the ITO surface. After rinsing and drying, the sample is ready for optical measurements.

*Polarization-Resolved Single-Molecule Localization-Based Microscopy*. This section describes the experimental approach that controls and studies the interaction between the single-molecule emitter and the plasmonic antenna with super resolution. We focus on three strategies to modulate and analyze the coupling between the fluorescent emitter and plasmonic nanoantenna: (i) modulate the polarization of the wide-field excitation laser; (ii) analyze the polarization of the QD emission pattern in the far-field; (iii) tune QD emission spectrum to be on- or off-plasmon resonance. As shown in Figure S2a, the modulation approach (i) is achieved by adding a polarizer in the excitation pathway and (ii) by adding an analyzer in the emission pathway of the wide-field fluorescence microscopy. Both the polarizer and analyzer are mounted on rotating stages (Figure S2b). As the polarizer/analyzer rotates, the far-field images of QDs are acquired at different excitation/emission polarization combinations. The excitation is filtered out, and the emission fluorescence is collected. No noticeable sample drift was observed during the measurement duration.

For a QD without plasmonic structures nearby, the fluorescence emission increases linearly with the excitation up to around half of the saturation intensity [22]. However, depending on the spectral overlap of the QD emission and plasmon resonance, the intensity of fluorescence emitted from the QD-plasmon system is no longer linear with the excitation and strongly affected by whether the QD emission is on- or off-resonance with the plasmon resonance mode [39]. More specifically, previous research has shown that, based on the emitter's location and emission spectra, the coupling strength between the emitter and the antenna fluctuates, causing shifts in the fitted location of the emitter [39]. As for the modulation approach (iii), we start with CdSe/ZnS QDs with an emission peak at around 800 nm. The results shown in the rest of this manuscript are generated based on this experimental setting. QDs with an off-resonance emission or dual emissions like Mn[+]-doped CdS QDs which have simultaneous on- and off-resonance emissions can be used to further investigate the effect of spectral overlap on the amount of emission coupled into the far-field via the plasmonic antenna and the shift in emitter localization.

For the polarization modulation experiment, we adopt a two-step measurement: (1) a measurement to index the apparent location of QDs together with extraction of local near-field intensity without the polarizer and analyzer, followed by (2) measurements with a modulated linearly polarized light at several different excitation/emission polarization angles. The imaging system with excitation only, emission only, and both polarization modulations together are first calibrated and characterized with sparsely spin-coated 800-nm emission QDs. Measurements are carried out on each individual unit composed of one antenna and one QD with the polarizer and the analyzer rotating separately from 0° to 360° with 45° intervals. For each excitation and emission polarization combination, a wide-field fluorescence image is captured, generating 81 images in total for one full circle rotation (Figure 3). Afterwards, the centroid position of each diffraction spot is determined. There are multiple methods for extracting the centroid position of the probe. One recently introduced is using a basis of Hermite–Gaussian functions as a PSF model to fit abnormal, multilobed PSFs generated from a system with dye labels close to plasmonic nanowires [29]. Here, the multilobed PSF is not observed in our QD-antenna system. Instead, we start with a typical 2D Gaussian function to fit the far-field image, and apply the maximum likelihood estimation method (MLEM) to optimize this PSF to give us the best fit, returning an estimation of QD's apparent position [40]. Even though the degree of final localization precision is heavily dependent upon the number of photons collected and the fluorescence background level, this fitting method has shown overall good performance for localizing QDs (or other emitting probes) in free space for both simulation and experimental results.

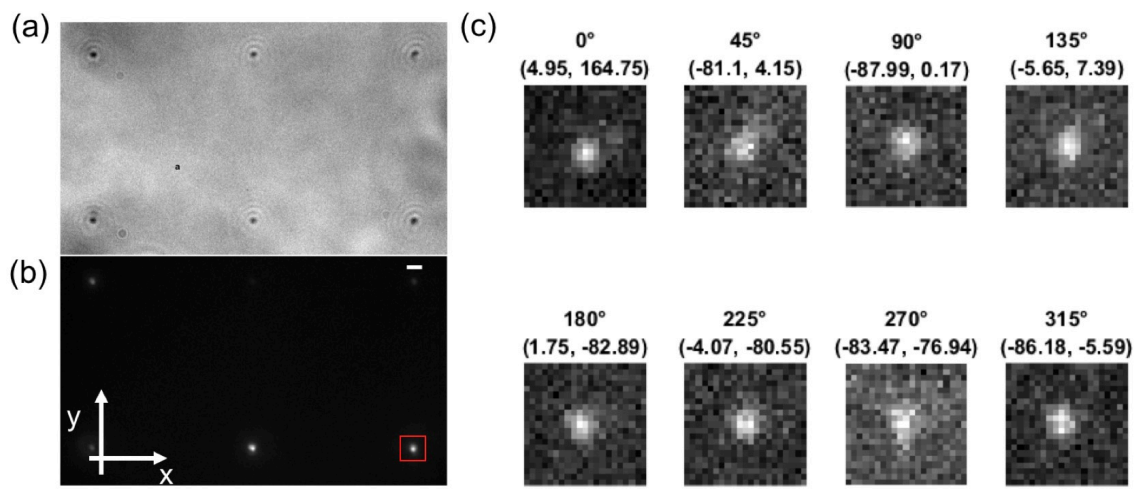

**Figure 3: QD images at different emission polarization**. (a) Bright-field image of nanostructure + QD. Since the size of the structure is below the diffraction limit, only the scattering light can be imaged without details of the structure. (b) Fluorescence image of QDs close to plasmonic structures. Different QD brightness may come from different coupling strength between the QD and the structure. Scale bar: 1 µm. (c) Images of single-molecule QD with different analyzer polarizations and displacements (labeled at the top of each figure, displacements in nm unit). Since QDs have degenerate excitation dipoles moments, changing the excitation polarization on "free" QDs would not affect the image. With the changing emission polarization, the center of the QD images shift.

When the emission is coupled into a plasmonic antenna, it is re-directed and re-radiated into the far-field compared to the radiation from the emitter alone. Due to the strong electromagnetic coupling of the emitter to the nearby plasmonic structure, the far-field radiation pattern of the emitter will be distorted, introducing imprecision to the localization of single emitters during this fitting step. The final QD far-field images are affected by a combination of interactions, like scattering from the nanostructure surface, dielectric distortion of the emission, and Young's interference effect between the emitter and the image dipole [30-31]. Thus, modulation and characterization of electromagnetic coupling strength between QDs and nearby metallic structures is crucial to understand and counteract mislocalization of QDs.

*Mislocalization of QDs close to plasmonic structures*. The fitted QD centers at different polarization combinations are compared to the center determined by the sum image of all polarizer and the analyzer polarization combinations. The spatial displacement is calculated as the difference between the two centers. The rotation of the polarizer in the excitation path does not show impact on the emitter displacement since QDs have degenerate excitation dipoles. This result verifies the result discussed in [29] that the excitation polarization does not affect the PSF distortion. However, as the analyzer rotates, we observe clear shifts in apparent QD center positions (mislocalizations) at different emission polarization. As shown in Figure 4a, displacement of the emitter in x (parallel) and y (perpendicular to the bottom arm of the L-shape) directions are plotted over one full circle of the analyzer rotation. To compensate for the discrepancy of photoblinking behaviors captured at different polarizations and its effect on the imaged fluorescence intensity, the final plot is the average over four full circles of the polarizer rotation. Photobleaching of QDs was negligible, and the few cases where it did occur were removed from the analysis pipeline. As expected, the shift of the emitter center positions in both x and y directions is periodic and repeats every 180° of the analyzer rotation (Figure 4). As the analyzer rotates, the shifts in the x- and y-axis vary and can be larger than 100 nm. After the optical measurement, the distance between the inner corner of the L-shaped nanoantenna and the QD is measured using SEM with the help of indexing markers patterned during the sample fabrication. This distance is regarded as the ground truth for the emitter's location and can be further used as the position of dipole source for the simulations (Figure 4b).

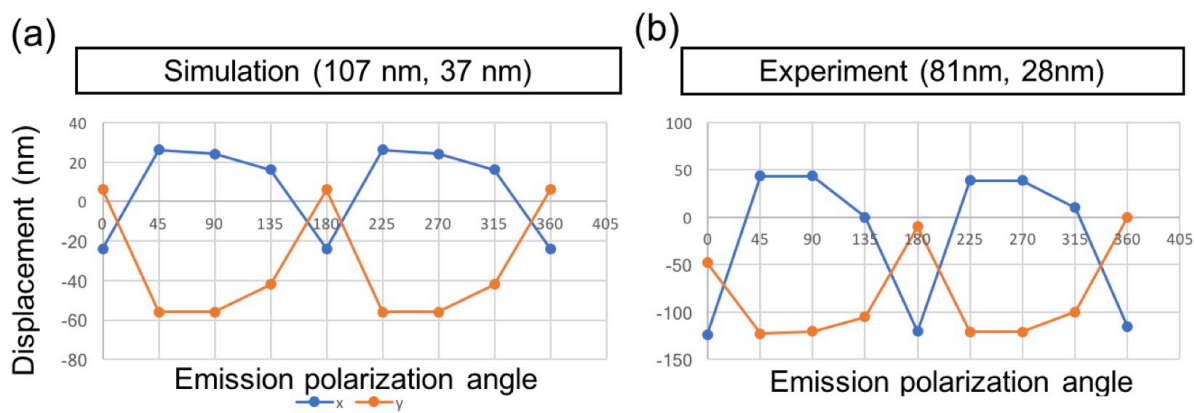

**Figure 4: Pattern matching to determine emitter's position**. Simulated (a) and experimentally measured (b) displacement are matched and the difference in the center location is the value of mislocalization. The change of mislocalization shares the same pattern as the simulation result. Based on the magnitude and direction of the displacement, the plots from the experiment can be compared with plots in the simulation collection, and patterns are matched using a least-squares model fit.

Next, a large set of FDTD simulations were performed to generate far-field images with different conditions for both on- and off-resonance emitters close to metallic systems. Based on the mislocalization trend as the emitter moving away from the inner corner of the antenna (Figure 5a-b), a simple qualitative model can be adopted to understand the interaction between the emitter and the plasmonic antenna. In addition, this model can also be used to determine how the interaction strength can influence the level of mislocalization for emitters at different locations.

For an L-shaped nanoantenna and QDs with on-resonance emission, the displacements are determined by superposition of dipole signals. Specifically, the system can be modeled as one quantum dot, which can be considered as a single dipole, interacting with two induced dipoles, each from one arm of the L-shaped metallic structure. The observed far-field image is formed by the superposition and interference of radiation from the three dipoles in total. The left arm mainly changes the displacement in the x-direction (perpendicular to the left arm), while the bottom arm changes the displacement in the y-direction. Compared to other one-dimensional structures studied in previous works, one significant difference for this system is that the emitter is coupled at two directions simultaneously with different coupling strengths. When the emitter is located on the symmetry axis of the L-shape, the interactions with the two arms are identical, and the displacements in the x- and y-directions are similar (Figure 5c). As the emitter moves asymmetrically away from one arm (for example, the left arm), the displacement range in the x-direction would first increase and then decrease. For the specific position shown in Figure 5c, when the emitter is away from the left arm (125 nm), its interaction with the left dipole is minimal so that its displacement in the x-direction is less than 20 nm. At the same time, its emission is still coupled to the bottom arm, making the maximum displacement in the y-direction to be almost 90 nm. Meanwhile, the oscillation of displacement as the emitter moving away from one arm shows that besides the superposition of radiation from multiple dipoles, interference between them also plays an essential role in determining the final displacement. Depending on the relative phases between three dipoles, a constructive/destructive interference may move the fitted center closer to/further away from the interface [30]. For instance, if the dipole (QD) is oriented parallel to the bottom arm, it interferes constructively with the induced dipole from the left arm and destructively with the one from the bottom arm. Thus, the mislocalization in the x-direction (0° emission polarization angle) fluctuates but stays negative for the measured distance range (Figure 4), which shares similar patterns as in Ref. [30]. Depending on how the QD is located and oriented in relation to the arm orientation, constructive or destructive interference may prevail.

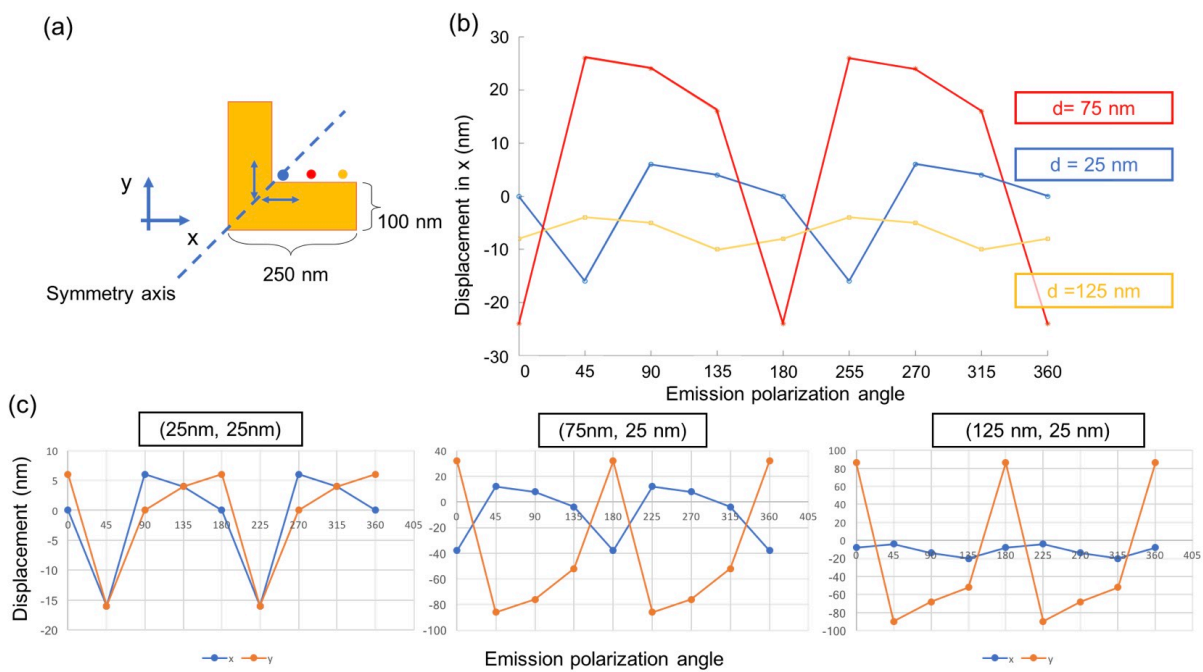

**Figure 5: Calculated displacement for emitters at different distance to the L-shaped nanoantennna**. (a) Cartoon demonstrating emitters at different locations that interact with two induced dipoles, each from one arm of the L-shape. The displacement in the x-direction for QDs positioned 25, 75 and 125 nm away from the left arm are plotted in (b). The colors of emitters correspond to plot colors. (c) Simulated mislocalizations of single-molecule QDs positioned at the inner corner with different distances to the L-shape. As the analyzer rotates from 0° to 360°, the shifts in x and y axis vary, and can be as large as 100 nm. The displacement is caused mainly by the super-position of the radiations from one original dipole plus two induced dipoles, as well as the interference between the radiations. Emitter at 75 nm has the most notable fluctuation and the greatest absolute displacement. When the emitter too close (25 nm), besides super-position, destructively interference reduces the displacement in the x-direction. When it is far away from the interface, the displacement decays.

Ideally, if the simulation and experimental settings are matched, the magnitude of the polarization-dependent PSF distortion and emitter displacement can be mapped out around the plasmonic nanoantenna. This mapping can be utilized to help deduce the actual emitter position from the distorted PSF by measuring the relative shifts. Meanwhile, the experimental results can be further fed back to simulations in order to improve the predictive power. In the end, the SEM step can be abandoned, and the real emitter position can be obtained purely from matching the experimental and simulation displacement plots. However, the sample preparation method requires high alignment precision between two lithography steps, and has proven to be challenging with respect to (1) nanostructures are fabricated with designed dimensions, (2) QDs are positioned close to nanostructures, (3) QDs stay fluorescent during the whole optical measurement, and (4) real locations of QDs are accessible using SEM. Currently, the 'post mortem' step with SEM to find the real position of the emitter has not been successful, so we are not able to include the actual position of the emitter in iterations that 'close the loop'. Instead, we generate a library of simulated images and plots, and match the displacement pattern from the experiment to the simulation result using a least-squares model fit. The dipole source position input from the simulation is then used as the 'real' position of the emitter when we see the change of displacement share the same pattern. A proposal for an alternative method for sample preparation using dip-pen nanolithography (DPN) is presented in the discussion section below.

*Detection of Plasmon-Induced Emission Polarization Rotation*. In addition to the substantial modification that QD-antenna interaction has on apparent QD localizations, we found that QD-nanoantenna interaction also strongly influences the polarization of the emitted light from this system (mispolarization). Previous studies have reported that the fluorescence emission rate and polarization can be re-directed depending on the design of the antenna (e.g., Yagi-Uda antenna or nanorod) [28,32]. The discussion below focuses on a more complex antenna shape, L-shape, and explores the change in polarization that results from plasmon-coupled emission. By measuring the emission polarization of a single-molecule QD coupled to an individual plasmonic nanoantenna, we reveal that QD's emission polarization can be significantly rotated depending on the emitter's position.

Far-filed images for QDs with different dipole orientations and emission wavelengths at various positions next to the antenna (H 60 nm, L 200 nm, W 50 nm) are calculated using the same simulation settings as described in previous sections. To quantify the rotation of the emission polarization, we define the apparent emission polarization ($\theta_{app}$) of each emitter by comparing the total intensity of images collected at two perpendicular directions

$$\theta_{app} = \arctan\sqrt{\frac{I_{\updownarrow}}{I_{\leftrightarrow}}}$$

where $I_{\leftrightarrow}$ and $I_{\updownarrow}$ are intensities collected at directions parallel and perpendicular to the bottom arm of the L-shape, respectively. The bottom arm and left arm are aligned to the x-axis and y-axis of a Cartesian coordinate system, respectively. Intensities are always positive and the arctangent function maps the apparent emission polarization into the first quadrant (0°~90°). The 2D projection of the emission polarization at the image plane can be collected with a polarization-resolved optical set-up shown in the previous section. Theoretically, a free single-molecule QD at the image plane can be considered as a dipole source, whose emission polarization is determined solely by its dipole orientation. However, when interacting with the plasmonic antenna nearby, a combination of near-field coupling, superposition, and interference with the nanoantenna far-field emission would introduce mislocalization and mispolarization, which hinders us from abstracting the actual plasmonic near-field intensity [24-27, 30-31]. Moreover, when a high numerical-aperture (NA) objective lens is used during image acquisitions, it has been reported that the cross-talk between different polarization channels would deviate the calculated $\theta_{app}$ [28]. This issue can be resolved by calibrating the imaging system using a control sample with free QDs randomly positioned on the coverslip for simulations or experiments. In this way, the relationship between the expected and the calculated $\theta_{app}$ can be mapped and used to correct the measurements.

Since the L-shaped metallic structure can be considered as two nanorods that are symmetrically connected, the rotation of the apparent polarization angle is dependent on the position and orientation of the emitter with respect to the symmetry axis. Table 1 shows a compilation of the calculated mispolarization of 800-nm emission QDs under different conditions. The large rotation for 0° and 90°-oriented dipoles is mainly contributed by the superposition of the induced localized surface plasmons (LSPs) modes. Radiation from the bottom arm increases $I_\leftrightarrow$, while radiation from the left arm increases $I_\updownarrow$. If the scattered radiation intensities from two arms are equal, the detected apparent emission polarization is expected to be 45°. For instance, for emitters positioned on the symmetry axis (e.g., (25 nm, 25nm) away from the inner corner of the L-shape, the interaction between the emitter and two arms is equal. Therefore, both 0° and 90°-oriented dipoles are rotated towards 45° while orientation of 45°-oriented dipoles are unchanged (Table 1 and Figure 6a). The degrees of rotation for 0° and 90°-orientations are symmetrical as expected. As the emitter moving away from both arms along the symmetry axis (e.g., (75 nm, 75 nm)), the induced radiation from both arms decreases, causing less rotation of the detected $\theta_{app}$ towards 45°. When the emitter is off the symmetry axis, depending on its distance to two arms, all three dipole orientations are 'mispolarized'. When QD moves away from the left arm (e.g., (75, 25 nm) and (125, 25 nm) in Figure 6), the interaction between them decreases. A higher $I_{\leftrightarrow}$ of the image is due to more dominant emission from the bottom arm. In this regard, $\theta_{app}$ rotates more towards 0° for both 0° and 90° dipole orientations. The total signal intensity collected at the 90° dipole orientation is also much higher (2x) than at 0°. Conversely, for QDs further away from the bottom arm, $\theta_{app}$ is closer to 90° because of the superposition of the induced LSP emission at the left arm with the original QD emission.

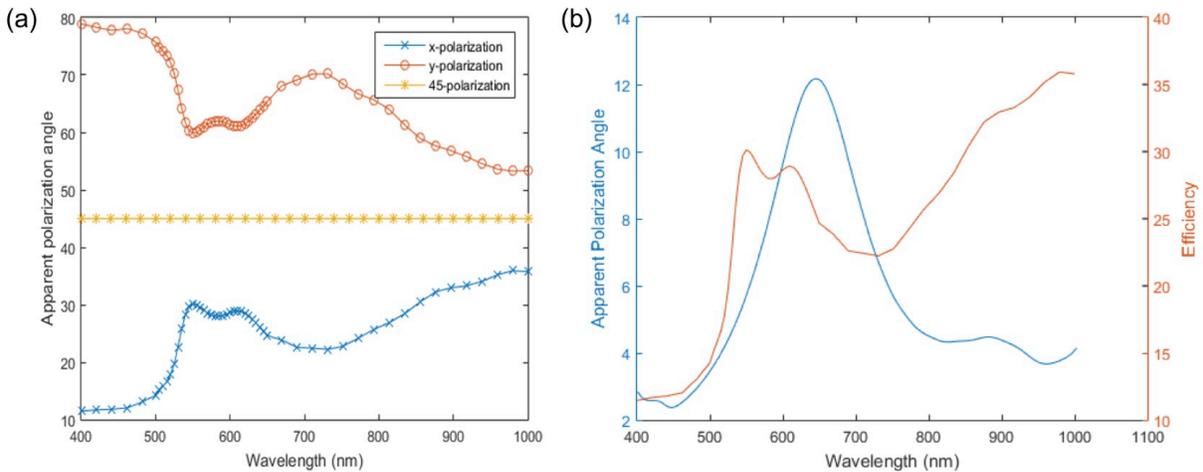

**Figure 6: Plasmon-induced rotation of the QD emission polarization**. (a) Calculated spectrum of rotation of apparent polarization angle for QD positioned at the inner corner with distance (25 nm, 25 nm) to the L-shape (L 200 nm, W 50 nm). For emitter positioned on the symmetry axis, since its interaction with two arms are identical, the apparent polarization is expected to rotate towards the axis direction. (b) Comparison of the $\theta_{app}$ (for the x-direction)

and plasmon resonance spectrum. The polarization angle spectrum shares similar pattern as the scattering cross-section under 800 nm, meaning that the scattering from the interfaces of two arms play an essential role in polarization rotation within the visible light spectrum.

**Table 1.** Calculated apparent emission polarization for different dipole orientations.

| Position (nm) / Dipole orientation (°) | (25, 25) | (75, 75) | (75, 25) | (125, 25) | (25, 75) | w/o antenna |
|---|---|---|---|---|---|---|
| 0 | 24.97 | 15.75 | 16.17 | 6.26 | 26.09 | 5.74 |
| 45 | 45.00 | 45.00 | 49.26 | 52.19 | 40.74 | 45.00 |
| 90 | 60.71 | 74.25 | 63.90 | 61.56 | 73.83 | 84.26 |

The similarity between the $\theta_{app}$ and the antenna scattering spectrum further verifies our hypothesis. In Figure 6a, the relationships between $\theta_{app}$ and emission wavelength of the emitter positioned at (25, 25 nm) are plotted for three different dipole orientations. Theoretically, since the emitter is positioned on the symmetry axis, it has identical types of interaction with both arms. Thus, the 0°- and 90°-oriented dipoles are mispolarized towards the symmetry axis while 45°-oriented keeps unchanged (Figure 6a). The greatest rotation is achieved at around the plasmon resonance peak, meaning that the on-resonance emission has a stronger scattering from both interfaces, rotating the apparent emission polarization more towards 45°. However, the difference between the spectrum of $\theta_{app}$ and scattering efficiency (Figure 6b), like double peaks and peak center shift, indicates that even though the scattering plays a major role in the mispolarization, there exist other factors that determines the final degree of rotation (e.g., interference between the QD and induced dipole emission). The two spectra diverge after around 800 nm, meaning that the same theoretical explanation might not be applicable for the infrared region. More investigation is needed to characterize other possible mechanisms.

In summary, we have demonstrated that the near-field interaction between a plasmonic structure and a nearby emitter is dependent on the emitter's relative position, emission polarization, and emission wavelength and can induce significant mislocalization and mispolarization during emitter's far-field detection and analysis. By isolating effects from the L-shaped nanoantenna's two arms at two orthogonal polarization directions, our study elucidates the mechanisms underlying modification of far-field emission polarization and localization in a QD-antenna system. Notably, we have confirmed that the strong interaction between the emitter and the nanoscale plasmonic structure can introduce substantial error to emitter localization. The apparent emission polarization can also measure the interaction strength. The emission polarization rotates towards the symmetry axis when the emitter is located on the axis. Depending on which arm emitter is closer to, the emission polarization rotates towards either the left or bottom arm of L-shape. By employing FDTD simulations and a polarization-resolved single-molecule localization-based method, we have revealed that the displacement of the emitter's localization originates mainly from the interference between the emitter and induced dipole emissions. In contrast, the superposition of the emitter and scattered radiation plays a more critical role in the emission polarization rotation. Even though we focused on a specific example, a similar analysis pipeline can be applied to study a more complex system (e.g., a metallic or dielectric nanoantenna whose shape contains sharp and rounded corners).

This topic can be further extended both in theory and application. In theory, an analytical model to quantitatively understand this QD-antenna system can be constructed and used to predict the true plasmonic near-fields. The effect of near-field coupling, far-field superposition, and interference can be implemented into the model and correct for the mislocalization and mispolarization. This model would provide a practical method that benefits many applications that rely on measuring field strengths with high precision, ranging from biology to high-speed integrated circuits to optical quantum computing. As for application, more simulated and polarization-modulated fluorescence images can be collected using other plasmonic structures, which help connect the emitter mislocalization at different emission polarizations to the shape of the plasmonic nanoantenna and emitter's distance to it. This information can be fed to a machine learning model (e.g., a convolutional neural network, CNN) to solve the inverse problem of predicting the plasmonic structure based on the far-field images. The L-shape will be one of the base structures for the model, and together with other base structures like nanowires and nanodisks, the whole contour of the structure can be determined by combining base structures.

**Methods**

**FDTD simulation and QD image generation**

Finite-difference time-domain (FDTD) simulations were performed using a free and open-source software package called Meep. Two types of simulations are carried out: (a) spectrum of scattering and absorption cross-sections of L-shaped nanoantennas using direct plane wave illuminations, and (b) electromagnetic near-field mapping of the QD-antenna system and image generation with a far-field projection. For all calculations, we assumed a background refractive index of 1 to mimic the dry sample condition in air. A glass substrate layer with a refraction index equal to 1.52 is positioned below the plasmonic nanoantenna. The dielectric function of gold was obtained from Johnson and Christy [42]. For (b), we consider QDs as dipole sources and simulated the radiation patterns of dipoles oriented perpendicular and parallel with respect to each arm of the L-shape and along the symmetry axis. A full range of QD distances to the L-shape inner corner and L-shape with different dimensions are also calculated. The dipole source orientation is assumed to be fixed under experimental conditions. A monitor is set up below the nanoantenna and QD to measure the electric fields of the near field. The collection angles that fall outside the NA of the objective lens are filtered out. The far-field projection and image generation are then calculated using a chirped-z transformation in MATLAB. The radiation of the isotropic emitter is calculated by adding together the images of the dipole source perpendicular and parallel to the antenna arm. This image can then be fitted with a 2D Gaussian function to determine the emitter's apparent location and compared with the input location of the simulation. All relevant data, codes for simulations and data analysis are available from authors on request.

**Sample fabrication**

Samples were fabricated on ITO coated glass coverslips (Nanocs). First, the coverslip is cleaned by a rinse with acetone, isopropanol and DI water in sequence. A thin film (~180 nm) of poly(methyl methacrylate) (PMMA) is then spin-coated onto the clean coverslip and baked in preparation for the pattern writing. L-shapes with various dimensions and markers are patterned into the photoresist using the electron beam lithography (first layer). The pattern is developed in a MIBK/IPA 1:3 solution, followed by a plasma etching step to create sharp edges and improve metal adhesion to the surface. A layer of gold (60 nm) is then deposited using an electron beam evaporator (CHA) with a wetting layer of titanium (1 nm). After the lift-off step in acetone, gold on areas without electron beam exposure are removed, leaving only L-shaped antennas on the coverslip. A similar sample preparation step is carried for the second layer of electron beam lithography. For this layer, the instrument can automatically align the markers of two layers and pattern a 200-nm square next to each L-shape. After removing the exposed photoresist, the ITO coverslip was soaked in the diluted QD solution (1 μm of Qdot™ 800 ITK™ Carboxyl Quantum Dots solution) for 60 min, followed by the photoresist lift-off, rinsing and drying.

**Optical measurements**

The samples were illuminated using a 642-nm laser diode source (Coherent) using an inverted microscope with a 100x oil-immersion objective (NA 1.49, Nikon). The excitation is filtered using a dichroic and an 800-nm long-pass filter. The excitation and emission polarizations are controlled through the polarizer and analyzer, respectively. Both the polarizer and analyzer are mounted on separate rotating stages controlled by Arduino, and QD images at different excitation/emission polarization combinations are recorded. Fluorescence was collected by an electron-multiplying charge-coupled device (EMCCD) camera with an extra set of lenses to increase the magnification further. The sample is mounted onto the sample stage and waited long enough until no apparent drift is observed. In each field-of-view (FOV), at least 25 QD-antenna units can be measured simultaneously. For each FOV, images are taken over four complete rotations of the polarizer (from 0° to 360° with 45° intervals). Later, the same FOV can be imaged using SEM with the help of patterned markers, and the true location of QDs can be compared with the measured locations.

**Dip-pen Nanolithography (DPN)**

As mentioned in previous sections, the current sample fabrication method requires precise alignment between two lithography steps to deposit QDs close to plasmonic structures, making the success rate reasonably low. A better approach that adopts the dip-pen nanolithography (DPN) method collaboratively with E-beam lithography has been carried out to directly 'write' single QDs with high precision (Figure S1). During DPN, an AFM tip first scans the sample to obtain the metallic landscape. The tip then moves away to dip into the QD ink, then moves directly back to the desired location for QD deposition. To guarantee single molecules rather than clusters of QDs, the concentration of QD solutions and dwell time during the print must be controlled [37,41]. Compared with our previous method, this scanning probe techniques replace the original random distribution of emitters with direct placement, leading to a much higher success rate. With the help of this approach, the relationship between QDs position, local plasmonic field intensity, and the mislocalization/mispolarization can be unveiled with an improved success rate and throughput. Thus, a better fundamental understanding of the coupling effects in plasmonic nanostructures can be constructed, which offers a promising route to improve the accuracy of near-field probing and sensing applications.

## AUTHOR INFORMATION

Y.M. and R.C.B. designed and conducted of the experiment pipeline. Y.M. analyzed the data and drafted the manuscript. A.B., V.C. and J. H. took charge of the DPN deposition and single-molecule confirmation measurement.

The authors declare no competing financial interests.

## ACKNOWLEDGMENT

This project is supported by National Science Foundation (NSF) Grant 1808766. Both the sample fabrication, optical measurement and simulations are carried out at University of California, Los Angeles (UCLA). DPN experiments are conducted at Los Alamos National Lab.

# Supporting Information.

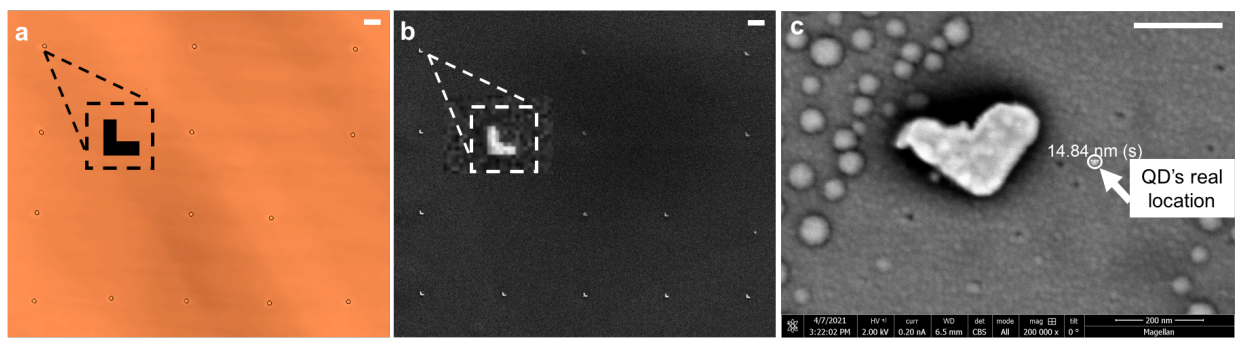

**Fig. S1: Single-molecule (SM) QD deposition using DPN.** (a) DPN printing spots indicated by brown circles. (b) L-shaped gold nanostructures under SEM. The pattern imaged with DPN tip and SEM matches. (c) Zoom-in of one L-shape with SM QD nearby. DPN first scan through the target area to determine the printing pattern. Then, the tip is inked and move to the start point with careful alignment. The ink with QDs is carried and printed by the tip (red circles in (a)) at locations determined by the first scan. The same target area is later moved under SEM (b), to quantify the relative positions of QDs and nanostructures (c). Scale bar: 200 nm.

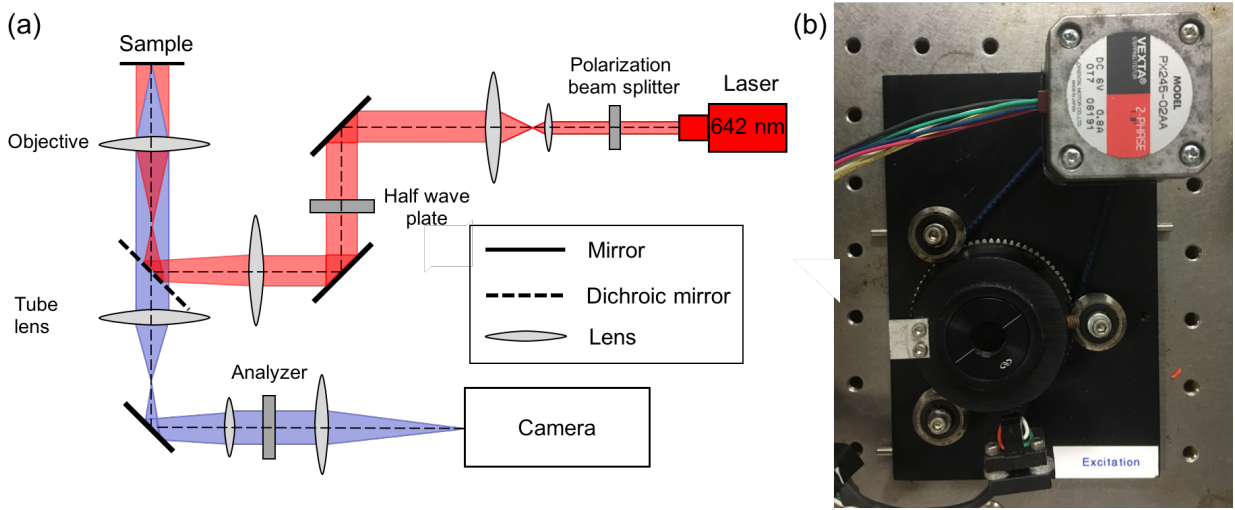

**Figure S2: Set-up for polarization-modulated super-resolution fluorescence imaging.** (a) Schematic diagram of the optical setup used for QD localization with excitation / emission polarization modulation. The half-wave plate in the excitation path works as a polarizer that rotates and controls the excitation polarization. The analyzer, on the other hand, selects a specific emission polarization. Both the polarizer and the analyzer are mounted on a rotating stage with 0.5° precision and controlled by the microcontroller Arduino. The stages are rotated between the optical measurements, and QD image at different excitation/emission polarization combinations are recorded. (b) The polarizer mounted on the rotating stage.